\documentclass{llncs}

\usepackage{breakurl} 
\usepackage{makeidx}
\usepackage{graphicx,latexsym,alltt}
\usepackage{amssymb,amsmath}
\usepackage{algorithm,algorithmic}
\usepackage{subfigure}
\usepackage{color}
\usepackage{url}
\usepackage{listings}
\usepackage{multirow} 
\usepackage{enumitem} 

\usepackage{multicol}

\newcommand{\figures}[1]{Figures/#1}
\newcommand{\biblio}[1]{../../../../../../Biblio/#1}

\definecolor{blank}{rgb}{0.55,0.55,0.55}

\long\def\comment#1{}

\renewcommand{\phi}{\varphi}

\def\defemb#1#2{\expandafter\def\csname #1\endcsname
                              {\relax\ifmmode #2\else\hbox{$#2$}\fi}}
\defemb{cA}{{\cal A}}
\defemb{cB}{{\cal B}}
\defemb{cC}{{\cal C}}
\defemb{cD}{{\cal D}}
\defemb{cE}{{\cal E}}
\defemb{cF}{{\cal F}}
\defemb{cG}{{\cal G}}
\defemb{cH}{{\cal H}}
\defemb{cI}{{\cal I}}
\defemb{cJ}{{\cal J}}
\defemb{cL}{{\cal L}}
\defemb{cM}{{\cal M}}
\defemb{cN}{{\cal N}}
\defemb{cO}{{\cal O}}
\defemb{cP}{{\cal P}}
\defemb{cR}{{\cal R}}
\defemb{cS}{{\cal S}}
\defemb{cT}{{\cal T}}
\defemb{cU}{{\cal U}}
\defemb{cV}{{\cal V}}
\defemb{cW}{{\cal W}}
\defemb{cX}{{\cal X}}
\defemb{cZ}{{\cal Z}}

\makeatletter
\newenvironment{prog}{\vspace{1.0ex}\par
\obeylines\@vobeyspaces\tt}{\vspace{1.0ex}\noindent
}
\makeatother
\newcommand{\startprog}{\begin{prog}}
\newcommand{\stopprog}{\end{prog}\noindent}

\catcode`\_=\active
\let_=\sb
\catcode`\_=12
\makeatother

\newif\ifpaperVersion
\paperVersiontrue
\ifpaperVersion
	\newcommand{\ignore}[1]{}
	\newcommand{\deleted}[1]{}
	
	\newcommand{\pending}[1]{}
	\newcommand{\done}[1]{}
	\newcommand{\doubt}[1]{}
	\newcommand{\josep}[1]{}
	\newcommand{\david}[1]{}
	\newcommand{\sergio}[1]{}
	\newcommand{\tama}[1]{}
\else
	\definecolor{ignoreColor}{rgb}{1,0.5,0}
	\definecolor{pendingColor}{rgb}{0.2,0.7,0.2}
	\definecolor{doneColor}{rgb}{0.7,0.2,0.7}
	\definecolor{doubtColor}{rgb}{0.6,0.6,0.4}
	\definecolor{josepColor}{rgb}{0.2,0.6,0.6}
	\definecolor{davidColor}{rgb}{0.6,0.2,0.6}
	\definecolor{sergioColor}{rgb}{1.0,0.5,0.0}
 	\definecolor{tamaColor}{rgb}{0.2,0.8,1.0}

	\newcommand{\ignore}[1]{\textcolor{ignoreColor}{\#Ignored: #1}}
	\newcommand{\deleted}[1]{\textcolor{red}{\#Deleted: #1}}
	
	\newcommand{\pending}[1]{\textcolor{pendingColor}{\#Pending: #1}}
	\newcommand{\done}[1]{\textcolor{doneColor}{\#Done: #1}}
	\newcommand{\doubt}[1]{\textcolor{doubtColor}{\#Doubt: #1}}
	\newcommand{\josep}[1]{\textcolor{josepColor}{\#JJJ: #1}}
	\newcommand{\david}[1]{\textcolor{davidColor}{\#DDD: #1}}
	\newcommand{\sergio}[1]{\textcolor{sergioColor}{\#SSS: #1}}
	\newcommand{\tama}[1]{\textcolor{tamaColor}{\#TTT: #1}}
\fi

\begin{document}

\frontmatter
\pagestyle{headings}
\addtocmark{Hamiltonian Mechanics}

\pagenumbering{arabic}

\title
{
 Erlang Code Evolution Control
\thanks
{
This work has been partially supported by MINECO/AEI/FEDER (EU)
under grants TIN2013-44742-C4-1-R and TIN2016-76843-C4-1-R,
and by the \emph{Generalitat Valenciana} under grant
PROMETEO-II/2015/013 (SmartLogic).
Sergio P\'erez was partially supported by the Spanish \emph{Iniciativa
de Empleo Juvenil} Programme (MINECO) and the European Social Fund in
collaboration with the \emph{Sistema Nacional de Garant\'{\i}a Juvenil} under
the grant PEJ-2014-A-24709 (\emph{Promoci\'on de Empleo Joven e Implantaci\'on
de la Garant\'{\i}a Juvenil 2014}, MINECO).
Salvador Tamarit was partially supported by the \emph{Conselleria de
Educaci\'on, Investigaci\'on, Cultura y Deporte de la Generalitat
Valenciana} under the grant APOSTD/2016/036.
}
}
\titlerunning{Erlang Code Evolution Control}

\author{David Insa\inst{1} \and Sergio P\'erez\inst{1} \and Josep Silva\inst{1} \and Salvador Tamarit\inst{1,2}}
\institute
{
Universitat Polit\`ecnica de Val\`encia\\
Cam\'i de Vera s/n, E-46022 Val\`encia, Spain\\
 \email{\{dinsa,serperu,jsilva,stamarit\}@dsic.upv.es}
\and
Universidad Polit\'ecnica de Madrid\\
Campus de Montegancedo s/n, E-28660 Boadilla del Monte, Spain\\
 \email{salvador.tamarit@upm.es}
}

\maketitle


\begin{abstract}
During the software lifecycle, a program can evolve several times for different reasons such as the optimisation of a bottle-neck, the refactoring of an obscure function, etc. 
These code changes often involve several functions or modules, so it can be difficult to know whether the correct behaviour of the previous releases has been preserved in the new release. Most developers rely on a previously defined test suite to check this behaviour preservation. We propose here an alternative approach to automatically obtain a test suite that specifically focusses on comparing the old and new versions of the code. Our test case generation is directed by a sophisticated combination of several already existing tools such as TypEr, CutEr, and PropEr;
and other ideas such as allowing the programmer to chose an expression of interest that must preserve the behaviour, or the recording of the sequences of values to which this expression is evaluated.  
All the presented work has been implemented in an open-source tool that is publicly available on GitHub. 
\end{abstract}


\section{Introduction}
\label{sec:intro}

\pending{Añadir una discusion (idealmente meterlo en la implementacion) sobre la posibilidad de generar inputs para los test cases a partir de la version evolucionada del sistema (y no solo desde la version inicial)}

During its useful life, a program might evolve many times. Each evolution is often composed of several changes that produce a new release of the software. There are multiple ways to control that these changes do not modify the intended behaviour of any critical part of the program. Most of the companies rely on \emph{regression testing} to assure that the desired behaviour of the original program is kept in the new release. There exist other alternatives such as the static inference of the impact of the changes \cite{jumpertz2010using,li2007testing,mongiovi2011safira,soares2013automated}. These techniques allow us to apply a change only when it is known that the behaviour will be preserved.

Even when a program is perfectly working and it fulfils all its functional requirements, sometimes we still need to improve parts of it. There are several reasons why a released program needs to be modified. For instance, improving the maintainability or efficiency; or for other reasons such as obfuscation, security improvement, parallelization, distribution, platform changes, and hardware changes, among others. 
%
%
Programmers that want to check whether the semantics of the original program remains unchanged in the new release usually create a test suite. 
There are several tools that can help in all this process. For instance, Travis CI can be easily integrated in a GitHub repository so that each time a pull request is performed, the test suite is launched.
We present here an alternative and complementary approach that allows for creating an automatic test suite to do regression testing. Our technique can check the evolution of the code even if no test suite has been defined.

In the context of debugging, programmers often use breakpoints to observe the values of an expression during an execution. Unfortunately, this feature is not currently available in testing, even though it would be useful to 
easily focus the test cases on one specific point without modifying the source code (as it happens when using asserts) or adding more code (as it happens in unit testing).
In this paper, we introduce the ability to specify \emph{points of interest} (POI) in the context of testing. A POI can be any expression in the code, e.g., a function call, meaning that we want to check the behaviour of that expression. 

In our technique, (1) the programmer identifies a POI, typically a variable\footnote{While our current implementation limits the POI to variables, nothing prevents the technique from accepting any expression as the POI.}, and a set of input functions, i.e., the starting points for the test cases. 
Then, by using a combination of random test case generation, mutation testing, and concolic testing, (2) the tool automatically generates a test suite that tries to cover all possible paths that reach the POI (trying also to produce execution paths that evaluate the POI several times). 
Therefore, in our setting, the \emph{input of a test case} (ITC) is defined as a call to an input function with some specific arguments, and the output is the sequence of those values the POI is evaluated to during the execution of the ITC. For the sake of disambiguation, in the rest of the paper we use the term \emph{traces} to refer to these sequences of values. 
Then, (3) the test suite is used to automatically check whether the behaviour of the program remains unchanged across new versions. This is done by passing each individual test case against the new version, and checking whether the same traces are produced at the POI. Finally, (4) the user is provided with a report about the success or failure of these test cases. Note that, as it is common in regression testing, this approach only works for deterministic executions. However, this does not mean that it cannot be used in a program with concurrency or other sources of indeterminism, it only depends on where the POI is placed.  

We have implemented our approach in a tool named \emph{SecEr}, which is publicly available at: \url{https://github.com/serperu/secer}. Instead of reinventing the wheel, some of the analyses performed by our tool are done by other existing tools such as CutEr~\cite{cuter}, a concolic testing tool, to generate an initial set of test cases that maximize the branching coverage; 
TypEr~\cite{typer}, a type inference system for Erlang, to obtain types for the input functions; and PropEr~\cite{proper}, a property-based testing tool, to obtain values of a given type. All the analyses performed by SecEr are transparent for the user. The only task that requires user inputs is identifying a suitable POI in both the old and the new version of the program. This task is easy when the performed changes are not too aggressive, but it could be more difficult when the similarities between both codes are hard to find. In those cases where the codes are too different, the returned expression of the main functions can be a good starting point to start checking the behaviour preservation. In case the user needs a more refined or intricate POIs, they could try to move them following backwards the control flows of both codes.

\begin{example}
Consider a real commit in the \texttt{string.erl} module of the standard library of the Erlang/OTP. The commit report is available at:
\begin{center}
\scriptsize
\url{https://github.com/erlang/otp/commit/53288b441ec721ce3bbdcc4ad65b75e11acc5e1b}
\end{center}

\noindent This change optimizes function \texttt{string:tokens/2}. We can automatically check whether this change preserves the original behaviour
with a single command of SecEr. We only need to indicate the two files that must be compared, and a POI for each file. Then, the tool automatically generates test cases that evaluate the POIs, trying to cover as many paths as possible. In this particular example, the POI is the output of function \texttt{string:tokens/2}. SecEr generated 6030 test cases that reached the POI, and it reported that both versions produce the same traces in the POI for all test cases.  
%

We can now consider a different scenario and introduce a simple error like, for instance, replacing the expression in line 248 of the new release with \texttt{tokens_single_2(S, Sep, Toks, Tok)}. In this scenario, SecEr generates 6010 test cases, but 497 of them have a mismatch in their traces. SecEr stores all the discrepancies found in a file, and it also reports one instance of these failing test cases: \texttt{tokens([11,5,9,1,22,3,9,4],[15])}, for which the original program produces the trace \texttt{[[11,5,9,1,22,3,9,4]]} while the new one produces the trace \texttt{[[4]]}.


\end{example}



\section{Overview of our approach to automated regression testing}
\label{sec:overview}

\begin{figure*}[h!]
\centering
\includegraphics[scale=0.32]{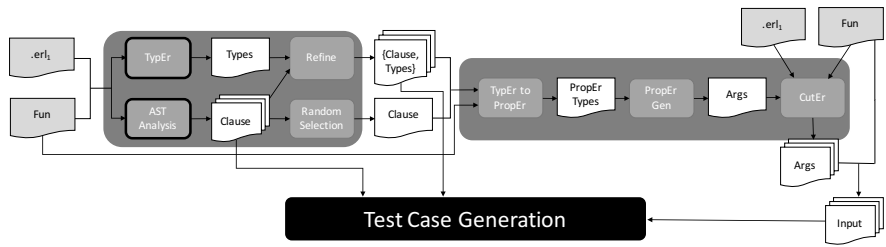}
\caption{Type analysis phase}
\label{fig:tap}
\end{figure*}

\begin{figure*}[h!]
\centering
\includegraphics[scale=0.4]{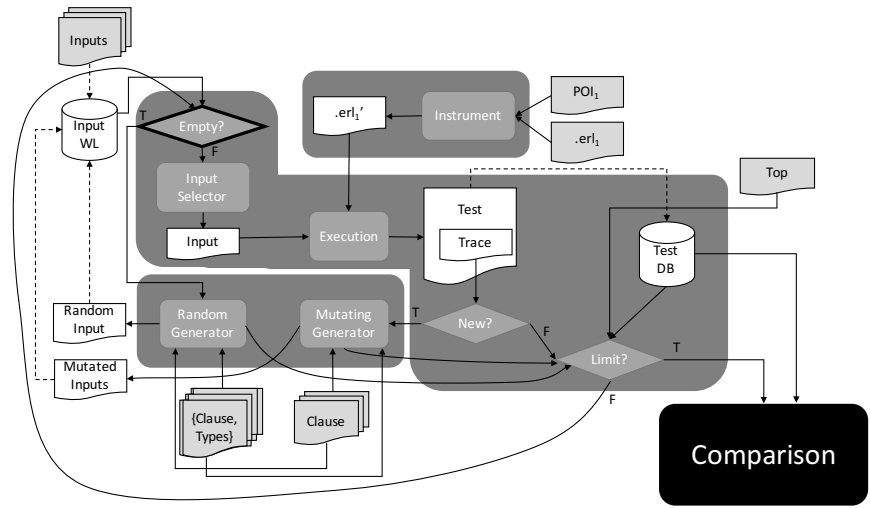}
\caption{Test case generation phase}
\label{fig:tgp}
\end{figure*}

\begin{figure*}[h!]
\centering
\includegraphics[scale=0.4]{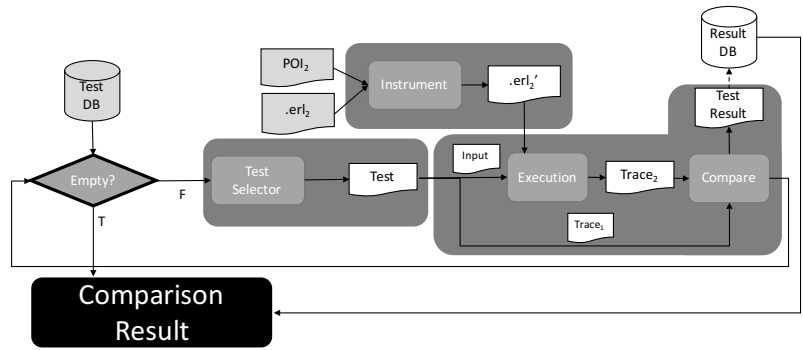}
\caption{Comparison phase}
\label{fig:cp}
\end{figure*}

Our technique is divided into three phases that are summarized in Figures~\ref{fig:tap}, \ref{fig:tgp}, and \ref{fig:cp}. In these figures, light grey boxes represent inputs, small dark grey boxes represent processes, white boxes represent intermediate results, and the big dark grey boxes are used to group several processes with a common objective. The initial processes of each phase are represented by a black-border box.

The first phase, depicted in Figure~\ref{fig:tap}, is a type analysis that is in charge of preparing all
 inputs 
 of the second phase (Test Case Generation). This phase starts by locating in the source code the Erlang module ({\tt .erl$_1$}) and a function ({\tt Fun}) specified in the user input (for instance, function \texttt{exp} in the \texttt{math} module). Then, TypEr is used to obtain the type of the parameters of that function. 
It is important to know that, in Erlang, a function is composed of clauses and, when a function is invoked, an internal algorithm traverses all the clauses in order to select the one that will be executed. 
Therefore, by analyzing the AST of the module, all the clauses of the input function are located. The types provided by TypEr are later refined to determine the type of each clause. All these clause types are used in the second phase, but in this phase we use PropEr to instantiate only one of them (e.g., \emph{$\langle$Number, Integer$\rangle$} can be instantiated to \emph{$\langle$4.22, 3$\rangle$} or \emph{$\langle$6, 5$\rangle$}). However, PropEr is unable to understand TypEr types, so a translation is previously performed. Finally, CutEr is fed with an initial call (e.g., \texttt{exp(4.22, 3)}) and provides a set of possible arguments (e.g., \{\emph{$\langle$1.5, 6$\rangle$}, \emph{$\langle$2, 1$\rangle$}, \emph{$\langle$1.33, 4$\rangle$}, $\dots$\}). Finally, this set is combined with the function to be called to generate the ITCs (e.g., \{\emph{exp(1.5, 6)}, \emph{exp(2, 1)}, \emph{exp(1.33, 4)}, $\dots$\}). All this process is explained in detail in Section~\ref{sec:type_dict}.

The second phase, shown in Figure~\ref{fig:tgp}, is in charge of generating the test suite. As an initial step, we instrument the program so that its execution records (as a side-effect) the sequence of values produced at the POI defined by the user. Then, we store all ITCs provided by the previous phase into a working list. Note that it is also possible that the previous phase is unable to provide any ITC due to the limitations of CutEr. In such a case, or when there are no more ITC left, we randomly generate a new one with PropEr and store it in the working list. Then, each ITC on the working list is processed by invoking it with the instrumented code. The execution provides us with the sequences of values the POI is evaluated to (i.e., the trace). This trace together with the ITC form a new test case, which is a new output of the phase. Moreover, whenever a not previously generated trace is computed, we mutate the ITC that generated that trace to generate more ITCs. The reason is that a mutation of this ITC will probably generate more ITCs that also evaluate the POI. This process is repeated until the specified limit of test cases is reached. All this process is explained in detail in Sections~\ref{sec:inst} and \ref{sec:test_gen}.

Finally, the last phase (shown in Figure~\ref{fig:cp}) checks whether the new version of the code passes the test suite. First, the source code of the new release is also instrumented to compute the traces produced at its POI. Then, all the generated test cases are executed and the traces produced are compared with the expected traces. 

\section{Detailed description of our approach}
\label{sec:detailed}

In this section, we describe in more detail the most relevant parts of our approach. We describe them in separate subsections.

\subsection{Initial ITC generation}
\label{sec:type_dict}


The process starts from the type inferred by TypEr for the whole input function. This is the first important step to obtain a significant result, because ITCs are generated with the types returned by this process, so the more accurate the types are, the more accurate the ITCs will be. The standard output of TypEr is an Erlang type specification returned as a string, which would need to be parsed. For this reason, we have hacked the Erlang module that implements this functionality to obtain the types in a data structure, easier to traverse and handle. In order to improve the accuracy, we define a type for each clause of the function ensuring that the later generated ITCs will match it. For this reason, TypEr types need to be refined to TypEr types per clause. 

However, the types returned by TypEr have  (in our context) two drawbacks that need to be corrected since they could yield to ITCs that do not match a desired input function. These drawbacks are due to the list length and due to the repeated variables. We explain both drawbacks with an example. Consider a function with a single clause whose header is \texttt{f(A,[A,B])}. For this function, TypEr infers the type \texttt{f( 1 | 2, [ 1 | 2 | 5 | 6, $\dots$ ] )}. 
Then, 
the type of the second parameter of the \texttt{f/2} function indicates that the feasible values for the second parameter are proper lists with a single constraint: it has to be formed with numbers from the set \texttt{[1,2,5,6]}. This means that we could build lists of any length, which is our first drawback. If we use these TypEr types, we may generate ITCs that will not match the function, e.g. \texttt{f(2,[1,1,5,6])}. 
On the other hand,  
the value relation generated by the repeated variable \texttt{A} is lost in the function type. This is due to the fact that the actual type of variable \texttt{A} is diluted in the type of the second argument. This could yield to mismatching ITCs if we generate, e.g., \texttt{f(1,[6,5])}. The last scenario exemplifies our second drawback.

Therefore, the types produced by TypEr are too imprecise in our context, because they may produce test cases that are useless (e.g., non-executable).
This problem is resolved in different steps of the process. In this step, we can only partially resolve the type conflict introduced by the repeated variables, such as the \texttt{A} variable in the previous example. The other drawback will be completely resolved during the ITC generation. To solve this problem, we traverse the parameters building a correspondence between each variable and the inferred TypEr type. Each time a variable appears more than once, we calculate its type as the intersection of both the TypEr type and the accumulated type. For instance, in the previous example we have \texttt{A = 1 | 2} for the first occurrence, and \texttt{A = 1 | 2 | 5 | 6} for the second one, obtaining the new accumulated type \texttt{A = 1 | 2}. 

Once we have our refined TypEr types, we rely on PropEr to obtain the input for CutEr. PropEr is a property-based testing framework with a lot of useful underlying functionality. 
One of them are the
term generators, which, given a PropEr type, are able to randomly generate terms belonging to such type. Thus, we can use the generators in our framework to generate values for a given type.

However, TypEr and PropEr use slightly different notations for their types, something reasonable given that their scopes are completely different. Unfortunately, there is not any available translator from TypEr types to PropEr types. In our technique, we need such a translator to link the inferred types to the PropEr generators. Therefore, we have built the translator by ourselves. Moreover, during the ITC generation, we need to deal with the previously postponed type drawbacks. For that, we use the parameters of the clause in conjunction with their types. To solve the first drawback, each time a list is found during the generation, we traverse its elements and generate a type for each element on the list. Thereby, we synthesize a new type for the list with exactly the same number of elements. The second drawback is solved by using a map from variables to their generated values. Each time a repeated variable is found we use the stored value instead of generating a new one.

We can feed CutEr with an initial call by using a randomly selected clause and the values generated by PropEr for this clause. CutEr is a concolic testing framework that generates a list of ITCs that tries to cover all the execution paths. Unfortunately, this list is only used internally by CutEr, so we have hacked CutEr to extract all these ITCs. Finally, by using this slightly modified version of CutEr we are able to generate the initial set of ITCs.


\subsection{Recording the traces of the point of interest}
\label{sec:inst}

There exist several tools available to trace Erlang executions \cite{redbug,dbgErlang,ttbErlang,erlyberly} (we describe some of them in Section~\ref{sec:rel}). However, none of them allows for defining a POI that points to any part of the code. Being able to trace any possible point of interest requires either a code instrumentation, a debugger, or a way to take control of the execution of Erlang. However, using a debugger (e.g., \cite{dbgErlang}) has the drawback that it does not provide a value for the POI when it is inside an expression whose evaluation fails. Therefore, we decided to instrument the code in such a way that, without modifying the semantics of the code, traces are collected as a side effect when executing the code.

The instrumentation process creates and collects the traces of the POI. To create the traces in an automatic way, we instrument the expression pointed by the POI. To collect the traces, we have several options. For instance, we can store the traces in a file and process it when the execution finishes, but this approach is inefficient. We follow an alternative approach based on message passing. We send messages to a server (which we call the tracing server) that is continuously listening for new traces until a message indicating the end of the evaluation is received. This approach is closer to the Erlang's philosophy. Additionally, it is more efficient since the messages are sent asynchronously resulting in an imperceptible overhead in the execution. As a result of the instrumenting process, the instrumented code sends to the tracing server the value of the POI each time it is evaluated, and the tracing server stores these values.

In the following, we explain in detail how the communication with the server is placed in the code. This is done by applying the following steps:

\begin{enumerate}
\item We first use the \texttt{erl_syntax_lib:annotate_bindings/2} function to annotate the AST of the code. This function annotates each node with two lists of variables: 
those variables that are being bound and those that were already bound in its subtree.

\item The next step is to find in the code the POI selected by the user. 
During the search process, we store the path followed in the AST with tuples of the form \texttt{(Node, ChildIndex)}, where \texttt{Node} is the AST node and \texttt{ChildIndex} is the index of the node in its parent's children array. Obtaining this path is essential for the next steps since it allows us to recursively update the tree in an easy and efficient way. When the POI is found, the traversal finishes. Thus, the output of this step is a path that yields directly to the POI in the AST.

\item Most of the times, the POI can be easily instrumented by adding a send command to communicate its value to the tracing server. However, when the POI is in the pattern of an expression, this expression needs a special treatment in the instrumentation. Let us show the problem with an example. 
Consider a POI inside a pattern {\tt \{1,POI,3\}}. If the execution tries to match it with {\tt \{2,2,3\}} nothing is send to the tracing server because the POI is never evaluated. Contrarily, if it tries to match it with {\tt \{1,2,4\}} we send the value {\tt 2} to the tracing server. Note that the matching fails in both cases, but due to the evaluation order, the POI is actually evaluated (and it succeeds) in the second case. 
We call target expression to the expressions that needs a special treatment in the instrumentation. In Erlang, these target expressions are: pattern-matching, list comprehensions and expressions with clauses (i.e., \texttt{case}, \texttt{if}, \texttt{functions}, \dots). The goal of this step is to divide the AST path into two sub-paths  \texttt{(PathBefore, PathAfter)}. \texttt{PathBefore} yields from the root to the deepest target expression (included), and \texttt{PathAfter} yields from the first children of the target expression to the POI.

\begin{figure}[!h]
\scalebox{0.9}{
\setlength\arraycolsep{5pt} 
$\begin{array}{ll}
\texttt{(LEFT\_PM)} & \texttt{p = e} \Rightarrow \texttt{p = ne}\\
~ & \mbox{if}~(\texttt{p = e},~\_) = last(PathBefore) ~\wedge~ pos_{path} = pos(\texttt{p}) \\
~ & \mbox{where}~(~(~\_~, ~pos_{path})~:~\_~) = PathAfter\\ 
~ & \hspace{0.5cm}\wedge~ \texttt{(\texttt{poi}, \texttt{npoi}, \texttt{np})} = pfv(\texttt{p}, PathAfter)\\
~ & \hspace{0.5cm}\wedge~ \texttt{ne} = 	\texttt{begin np = e, tracer!\{add, npoi\}, np end}\\
\texttt{(PAT\_GEN\_LC)} & \texttt{[e || gg]} \Rightarrow \texttt{[e || ngg]}\\
~ & \mbox{if}~(\_,\texttt{[e || gg]}) = last(PathBefore)\\
~ & \hspace{0.5cm}~\wedge~ \exists~ i. ~ 1 ~\leq~i ~\leq~length(\texttt{gg})\\
~ & \hspace{1cm}~\mbox{s.t.}~ \texttt{p_{gen} <- e_{gen}} = \texttt{gg}_i  ~\wedge~ pos_{path} = pos(\texttt{p_{gen}})\\
~ & \mbox{where}~(~\_~:~path_{tail}~)= PathAfter ~\wedge~ ((~\_~,~ pos_{path}~)~:~\_~)= path_{tail}\\ 
~ & \hspace{0.5cm}~\wedge~(\texttt{poi}, \texttt{npoi}, \texttt{np_{gen}}) =  pfv(\texttt{p_gen},~ path_{tail})\\
~ & \hspace{0.5cm}~\wedge~\texttt{ng =  p_gen <- begin tracer!\{add, npoi\}, [np_gen] end}\\
~ & \hspace{0.5cm}~\wedge~\texttt{ngg} =  \texttt{gg}_1 \ldots \texttt{gg}_{i-1}, \texttt{np_{gen} <- e_{gen}},~\texttt{ng},~ \texttt{gg}_{i+1} \ldots \texttt{gg}_{length(\texttt{gg})}\\
\texttt{(CLAUSE\_PAT)} & \texttt{e} \Rightarrow \texttt{ne}\\
~ & \mbox{if}~(\_,\texttt{e}) = last(PathBefore) ~\wedge~ \texttt{cls_e} = clauses(\texttt{e}) \neq \emptyset\\
~ & \hspace{0.5cm}~\wedge~ \exists~ i. ~ 1 ~\leq~i ~\leq~length(\texttt{cls_e})\\
~ & \hspace{1cm}~\mbox{s.t.}~ \texttt{p_{c} when g_{c} -> b_{c}} = \texttt{cls_e}_i  ~\wedge~ pos_{path} = pos(\texttt{p_{c}})\\
~ & \mbox{where}~(~\_~:~path_{tail}~)= PathAfter ~\wedge~ ((~\_~,~ pos_{path}~)~:~\_~)= path_{tail}\\ 
~ & \hspace{0.5cm}~\wedge~(\texttt{poi},~\texttt{npi},~ \texttt{np_{c}}) =  pfv(\texttt{p_c},~ path_{tail})\\
~ & \hspace{0.5cm}~\wedge~\texttt{nb_c = begin tracer!\{add, npoi\}, case np_{c} of \texttt{cls_e} end end}\\
~ & \hspace{0.5cm}~\wedge~\texttt{ncl} =\texttt{np_{c} when true -> nb_{c}}\\
~ & \hspace{0.5cm}~\wedge~\texttt{ncls_e} = \texttt{cls_e}_i,\ldots,\texttt{cls_e}_{i - 1}, \texttt{ncl}, \texttt{cls_e}_{i + 1},\ldots,\texttt{cls_e}_{length(\texttt{cls_e})}\\
~ & \hspace{0.5cm}~\wedge~\texttt{ne} = change\_clauses(\texttt{e},~ \texttt{ncls_e})\\
\texttt{(CLAUSE\_GUARD)} & \texttt{e} \Rightarrow \texttt{ne}\\
~ & \mbox{if}~(~\_~,\texttt{e}~) = last(PathBefore) ~\wedge~ \texttt{cls_e} = clauses(\texttt{e}) \neq \emptyset\\
~ & \hspace{0.5cm}~\wedge~ \exists~ i. ~ 1 ~\leq~i ~\leq~length(\texttt{cls_e})\\
~ & \hspace{1cm}~\mbox{s.t.}~ \texttt{p_{c} when g_{c} -> b_{c}} = \texttt{cls_e}_i  ~\wedge~ pos_{path} = pos(\texttt{g_{c}})\\
~ & \mbox{where}~(~\_~:~path_{tail}~)= PathAfter ~\wedge~ ((~\_~,~ pos_{path}~)~:~\_~)= path_{tail}\\ 
~ & \hspace{0.5cm}~\wedge~(\texttt{poi},~\_~) = last(PathAfter)\\
~ & \hspace{0.5cm}~\wedge~\texttt{nb_c = begin tracer!\{add, poi\}, case np_{c} of \texttt{cls_e} end end}\\
~ & \hspace{0.5cm}~\wedge~\texttt{ncl} =\texttt{p_{c} when true -> nb_{c}}\\
~ & \hspace{0.5cm}~\wedge~\texttt{ncls_e} = \texttt{cls_e}_i,\ldots,\texttt{cls_e}_{i - 1}, \texttt{ncl}, \texttt{cls_e}_{i + 1},\ldots,\texttt{cls_e}_{length(\texttt{cls_e})}\\
~ & \hspace{0.5cm}~\wedge~\texttt{ne} = change\_clauses(\texttt{e},~ \texttt{ncls_e})\\
\texttt{(EXPR)} & \texttt{poi} \Rightarrow \texttt{begin tracer!\{add, poi\}, poi end}\\
~ & \mbox{otherwise, where}~(~\texttt{poi},~\_~) = last(PathAfter)\\ 
\end{array}$}
\caption{Instrumention rules for tracing.}
\label{fig:trans_rules}
\end{figure}

\begin{figure}[!t]
$ pfv(p,~path) = $\\
$ ~~~~~~~~~~\left\{\setlength\arraycolsep{5pt} 
\begin{array}{ll} 
(poi,~ poi',~ p'') & \mbox{if}~ path = [(poi, pos)]\\
~& \hspace{-1cm}\mbox{where}~poi' = fv() ~\wedge~ p' = fv\_from(pos,~p) \\
~& \hspace{-1cm}\wedge~ p'' = p'_1\ldots p'_{pos-1}, poi',p'_{pos + 1}\ldots p'_{length(p)} \\ 
(poi,~ poi',~ p''') & \mbox{if}~ path = ((\_, pos):path_{tail}) \\
~& \hspace{-1cm}\mbox{where}~ p' = fv\_from(pos,~p) ~\wedge~ (poi, poi', p'') = pfv(p'_{pos}, path_{tail})\\
~& \hspace{-1cm}\wedge~ p''' = p'_1\ldots p'_{pos-1}, p'',p'_{pos + 1}\ldots p'_{length(p)} \\ 
\end{array} \right.$
%
~\\
~\\
$ fv\_from(pos,~p) =$\\
$~~~~~~p'_1\ldots p'_{pos},fv()_{pos+1} \ldots fv()_{length(p)}~\mbox{where}~(p'_1\ldots p'_{pos}, \_) = cv(p_1\ldots p_{pos}, ~[])$\\
~\\
$cv(list, map)$\\
$ ~~~~~~~~~~\left\{\setlength\arraycolsep{5pt} 
\begin{array}{ll} 
([], map) & \mbox{if}~ list = []\\
((fv:p_t'), map') & \mbox{if}~ list = (p_h:p_t) \wedge is\_var(p_h) \wedge \neg~ is\_bound(p_h) \\
~& \hspace{-1.5cm}\mbox{where}~ fv = fv() ~\wedge~ (p_t',~ map') = cv(p_t, map \cup \{p_h \mapsto fv\})\\
((fv_{map}:p_t'), map') & \mbox{if}~ list = (p_h:p_t) \wedge is\_var(p_h) \wedge p_h \mapsto fv_{map} \in map \\
~& \hspace{-1.5cm}\mbox{where}~ (p_t', map') = cv(p_t, map)\\
((p_h':p_t'), map'') & \mbox{otherwise} \\
~& \hspace{-1.5cm}\mbox{where}~ (p_h:p_t) = list~\wedge~ (children'_{p_h}, map') = cv(children(p_h), map)\\
~&\hspace{-0.85cm} \wedge~ p_h' = change\_children(p_h, children'_{p_h})\\
~&\hspace{-0.85cm} \wedge~ (p_t', map'') = cv(p_t, map')\\
\end{array} \right.$
\caption{Function $pfv$}
\label{fig:fvfun}
\end{figure}

%



\item Finally, the last step is the one in charge of performing the actual instrumentation. The \texttt{PathBefore} path is used to traverse the tree until the deepest target expression that contains the POI is reached. At this point, five rules (described below) are used to transform the code by using \texttt{PathAfter}. Finally, the \texttt{PathBefore} is traversed backwards to update the AST of the targeted function. 
The five rules are depicted in Figure \ref{fig:trans_rules}. 
The first four rules are mutually exclusive, and when none of them can be applied, the rule \texttt{(EXPR)} is applied. Rule \texttt{(LEFT\_PM)} is fired when the POI is in the pattern of a match expression. Rule \texttt{(PAT\_GEN\_LC)} is used to transform a list comprehension when the POI is in the pattern of a generator. Finally, rules \texttt{(CLAUSE\_PAT)}\footnote{Function clauses needs an additional transformation that consists in storing all the parameters inside a tuple so that they could be used in case expressions.} and \texttt{(CLAUSE\_GUARD)} transform an expression with clauses when the POI is in the pattern or in the guard of one of its clauses, respectively.
There are several functions used in the rules that need to be introduced. Function $last(list)$ returns the last element of a $list$. Function $pos(e)$ returns the child index of an expression $e$.  Function $is\_bound(e)$ returns {\tt true} if $e$ is bounded according to the AST binding annotations (see step 1). Function $clauses(e)$ and $change\_clauses(e, clauses)$ obtains and modifies the clauses of $e$, respectively. Finally, there is a key function named $\mathit{pfv}$, introduced in Figure~\ref{fig:fvfun}, that transforms a pattern so that the constraints after the POI do not inhibit the sending call. This is done by replacing all the terms on the right of the POI with free variables that are built using $\mathit{fv}$ function. Unbound variables on the left and also in the POI are replaced by fresh variables to avoid the shadowing of the original variables. In the $\mathit{pfv}$ function, $children(e)$ and $change\_children(e, children)$ are used to obtain and modify the children of expression $e$, respectively. 
\end{enumerate}

\subsection{Generation of new test cases using PropEr and test mutation}
\label{sec:test_gen}


The test case generation phase uses CutEr because its concolic analyses tries to generate 100\% branch coverage test cases. However, sometimes these analyses require too much time and we have to abort its execution. This means that after executing CutEr, we might have no test cases. Moreover, CutEr is not exhaustive enough when it evaluates a single branch. Therefore, in our context, it is insufficient to detect behaviour differences in a concrete branch, e.g. a wrong operator. 

Therefore, 
we should produce more test cases to increase the reliability of the test suite. One option is to use the PropEr generator to randomly synthesize new test cases, but this would produce many useless test cases (e.g., test cases that do not execute the POI). Hence, in order to avoid a completely random test case generation, we use a test mutation technique. The function that generates the test cases is depicted in Figure~\ref{fig:tgfun}. 
The result of the function is a map from the different obtained traces to the set of ITCs that produce them.
The first call to this function is $tgen(top, cuter\_tests, \emptyset)$, where $top$ is a user-defined limit of the desired number of test cases\footnote{In SecEr, a timeout is also used as a way to stop the test case generation.} and $cuter\_tests$ are the test cases that CutEr generates. Function $tgen$ uses the auxiliary functions $proper\_gen$, $trace$, and $mut$. The function $proper\_gen()$ simply calls PropEr to generate a new test case, while function $trace(input)$ obtains the corresponding trace when the ITC $input$ is executed. The size of a map, $size(map)$, is the total amount of elements stored in all lists that belong to the map. Finally, function $mut(input)$ obtains a set of mutations for the ITC $input$, where, for each argument in $input$, a new test case is generated by replacing the argument with a randomly generated value (using PropEr) and leaving the rest of arguments unchanged. 

\begin{figure}[!t]
$ tgen(top, pending, map) = $\\
$ ~~~~~~~~~~\left\{\setlength\arraycolsep{5pt} 
\begin{array}{ll} 
map & \mbox{if}~ size(map) \geq top\\
tgen(top, pending', map') & \mbox{if}~ size(map) < top\\
& \wedge~\exists~ input \in pending\\
& ~~~\mbox{s.t.}~ trace(input) \mapsto \_ \not\in map\\
 & \hspace{-3cm}\mbox{where}~ pending' =(pending \cup mut(input)) \backslash \{input\} \\
 & \hspace{-2.4cm} \wedge~ map' = map \cup \{trace(input) \mapsto \{input\}\}\\
 tgen(top, \{proper\_gen()\}, map') & \mbox{if}~ size(map) < top\\
& \wedge~\not\exists~ input \in pending\\
& ~~~~\mbox{s.t.}~ trace(input) \mapsto \_ \not\in map\\
 & \hspace{-3cm}\mbox{where}~ map' = map\\
 & \hspace{-2cm} \cup~ \{trace(input_{p}) \mapsto( \{input_{p}\} \cup inputs_{tp})\\
 & \hspace{-1,55cm}  ~|~ input_{p} \in pending \wedge trace(input_{p}) \mapsto inputs_{tp} \in~ map\}
\end{array} \right.$
\caption{Test generation function}
\label{fig:tgfun}
\end{figure}

\section{The SecEr Tool}
\label{sec:tool}

In this section we first briefly describe SecEr, how it is used and the parameters that it needs to automatically obtain test cases from a source code. Then, we provide a use case to exemplify how SecEr can be used to check behavioural changes in the code.

\subsection{Tool description}
SecEr is able to automatically generate a test suite that checks the behaviour of a POI given two versions of the same program. There are two ways of running this tool. The first one allows for generating and storing a test suite of a program that is specific to test a given POI. The second one compares the traces of two different versions of a program and reports the discrepancies.

\begin{lstlisting}[basicstyle=\ttfamily\scriptsize, frame=single, caption={\texttt{SecEr} command format}, label=lst:command]
$ ./secer -f FILE -li LINE -var VARIABLE [-oc OCCURRENCE] 
         [-f FILE -li LINE -var VARIABLE [-oc OCCURRENCE]]
         [-funs INPUT_FUNCTIONS] -to TIMEOUT
\end{lstlisting}

Listing \ref{lst:command} shows the usage of the SecEr command. If we want to run the command to only generate a test suite, we need to provide the path of the target file ({\small\texttt{FILE}}), the POI ({\small\texttt {LINE,VARIABLE}}, and {\small\texttt {OCCURRENCE}}), a list of initial functions ({\small\texttt{INPUT\_FUNCTIONS}}),\footnote{The format for this list is {\small\texttt{[FUN1/ARITY1, FUN2/ARITY2 ...]}}. If the user does not provide it, all functions exported by the module are used as input functions.} and a timeout ({\small\texttt{TIMEOUT}}). On the other hand, if we want to perform a comparison, we also need to provide the path of the second file and the POI to be compared. 

Note that, in the implementation, the limit used to stop generating test cases is a timeout, while the formalization of the technique uses a number to specify the amount of test cases that must be generated (see variable $top$ in Section~\ref{sec:test_gen}). 
This is not a limitation, but a design decision to increase the usability of the tool. The user cannot know a priory how much time it could take to generate an arbitrary number of test cases. Hence, to make the tool predictable and give the user control over the computation time, we use a timeout. Thus, SecEr generates as many test cases as the specified timeout permits. 

\subsection{Use case}
In order to show the potential of the tool, we provide an example to compare two versions of an Erlang program that computes happy numbers. They are taken from the Rosetta Code repository:
\begin{center}
\footnotesize
\url{http://rosettacode.org/wiki/Happy_numbers#Erlang}.
\end{center}

In order to unify their interfaces, we have made small changes. In concrete, in the {\tt happy0} module (Listing \ref{lst:happy0}) we have replaced {\small\texttt{main/0}} with {\small\texttt{main/2}} making it applicable for a more general case. In the {\tt happy1} module (Listing \ref{lst:happy1}) we have added the {\small\texttt{Happy}} variable (line 18), which stores the result of {\small\texttt{is_happy(X,[])}}. In both modules, we have added a type specification (represented with {\tt spec} in Erlang) in order to obtain more representative test cases. 

\begin{multicols}{2}
\begin{lstlisting}[tabsize=2,basicstyle=\ttfamily\tiny, frame=single, caption={happy0.erl}, label=lst:happy0]
1  -spec main(pos_integer(),pos_integer()) -> 
2  	[pos_integer()].
3  main(N, M) -> 
4  	happy_list(N, M, []).
5  
6  happy_list(_, N, L) when length(L) =:= N -> 
7  	lists:reverse(L);
8  happy_list(X, N, L) -> 
9  	Happy = is_happy(X),
10 	if Happy -> 
11 		happy_list(X + 1, N, [X|L]);
12 	true -> 
13 		happy_list(X + 1, N, L) end.
14 
15 is_happy(1) -> true;
16 is_happy(4) -> false;
17 is_happy(N) when N > 0 ->
18 	N_As_Digits = 
19 		[Y - 48 || 
20 		Y <- integer_to_list(N)],
21 	is_happy(
22 		lists:foldl(
23 			fun(X, Sum) -> 
24 				(X * X) + Sum 
25 			end, 
26 			0, 
27 			N_As_Digits));
28 is_happy(_) -> false.
\end{lstlisting}
\columnbreak
\begin{lstlisting}[tabsize=2,basicstyle=\ttfamily\tiny, frame=single, caption={happy1.erl}, label=lst:happy1]
1  is_happy(X, XS) ->
2  	if
3  		X == 1 -> true;
4  		X < 1 -> false;
5  		true ->
6  			case member(X, XS) of
7  				true -> false;
8  				false ->
9  					is_happy(sum(map(fun(Z) -> Z*Z end, 
10 						[Y - 48 || Y <- integer_to_list(X)])),
11 					  [X|XS])
12 			end
13 	end.
14 happy(X, Top, XS) ->
15 	if
16 		length(XS) == Top -> sort(XS);
17 		true ->
18 			Happy = is_happy(X,[]),
19 			case Happy of
20 				true -> happy(X + 1, Top, [X|XS]);
21 				false -> happy(X + 1,Top, XS)
22 			end
23 	end.
24 	
25 -spec main(pos_integer(),pos_integer()) -> 
26 	[pos_integer()].
27 main(N, M) -> 
28 	happy(N, M, []).
\end{lstlisting}
\end{multicols}
\vspace{-2mm}
Listing \ref{lst:correct} shows the execution of SecEr when comparing both implementations of the program with a timeout of 15 seconds. The POI is the {\tt Happy} variable, which is evaluated many times per execution, returning a multiple-valued trace. As we can see, the execution of both implementations behaves identically with respect to the Happy variable in the 320 generated test cases. 

\begin{lstlisting}[basicstyle=\ttfamily\scriptsize, frame=single, caption={\texttt{SecEr} reports that no discrepancies exist}, label=lst:correct]
$ ./secer -f happy0.erl -li 9  -var Happy -oc 1 
          -f happy1.erl -li 18 -var Happy -oc 1 
          -fun [main/2] -to 15
          
Function: main/2
----------------------------
Generated test cases: 320
Both versions of the program generate identical traces for the 
point of interest
----------------------------
\end{lstlisting}

In order to see the output of the tool when the behaviour of the two compared programs differ, we have introduced two different errors in the {\tt happy1} module. The first error is introduced by replacing the whole line 4 with  {\small\texttt{X < 10 -> false;}}. 
With this change the behaviour of both programs is different, and this is detected by SecEr, which produces the output shown in Listing \ref{lst:errorlength}. 
The second error consists in the replacement of the whole line 21 with {\small\texttt{false -> happy(X + 2, Top, XS)}}. 
The output of SecEr in this case is depicted in Listing \ref{lst:errorvalues}.
As we can observe in both errors, it does not matter where the bug is located as long as the bug affects the values computed at the POI.  Another interesting feature of the tool arises when we have a POI that is evaluated several times during the execution of the program. In this case, SecEr allows us to differentiate between two kinds of errors: Traces that differ in their number of elements (Listing \ref{lst:errorlength}) and traces with the same number of elements but with different values or different order (Listing \ref{lst:errorvalues}).

\begin{lstlisting}[basicstyle=\ttfamily\scriptsize, frame=single, caption={\texttt{SecEr} reports discrepancies (different trace length)}, label=lst:errorlength]
Function: main/2
----------------------------
Generated test cases: 251
Mismatching test cases: 22 (8.76%)
All mismatching results were saved at: ./results/main_2.txt
--- First error detected ---
Call: main(4,1)
happy0 trace (9,Happy,1): [false,false,false,true]

happy1 trace (21,Happy,1): [false,false,false,false,false,false,true]
\end{lstlisting}

\begin{lstlisting}[float,floatplacement=H,basicstyle=\ttfamily\scriptsize, frame=single, caption={\texttt{SecEr} reports discrepancies (different trace values)}, label=lst:errorvalues]
Function: main/2
----------------------------
Generated test cases: 289
Mismatching test cases: 263 (91.0%)
All mismatching results were saved at: ./results/main_2.txt
--- First error detected ---
Call: main(1,7)
happy0 trace (9,Happy,1): [true,false,false,false,false,false,true,false,
false,true,false,false,true,false,false,false,false,false,true,false,
false,false,true,false,false,false,false,true]
 
happy1 trace (21,Happy,1): [true,false,false,false,false,true,false,true,
false,false,false,false,false,false,false,true,false,true,true,false,
false,false,false,false,false,false,false,true]

\end{lstlisting}

\section{Related work}
\label{sec:rel}

Automated behavioral testing techniques like Soares et al. \cite{soares2013automated} and Mongiovi \cite{mongiovi2011safira} are very similar to our approach, but their techniques are restricted in the kind of changes that can be analyzed (they only focus on refactoring). 
Contrarily, our approach is independent of the kind (or the cause) of the changes, being able to analyze the effects of any change in the code regardless of its structure. 

Automated regression test generation techniques like Korel and Al-Yami \cite{korel1998automated} are also very similar to our approach, but they can only generate test cases if they have available both the old and the new releases. Contrarily, in our approach we can generate tests with a single release, and reuse the test cases to analyze any new releases by only specifying the points of interest.

Yu et al. \cite{yu2012practical} presented an approach that combines coverage analysis and delta debugging to locate the sources of the regression faults introduced during some software evolution. Their approach is based on the extraction and analysis of traces. Our approach is also based on traces although not only the goals but also the inputs of this process are slightly different. In particular, we do not require the existence of a test suite (it is automatically generated), while they look for the error sources using a previously defined test suite. 
Similarly, Zhang et al. \cite{zhang2013injecting} use mutation injection and classification to identify commits that introduce faults. 

Most of the efforts in regression testing research have been put in the regression testing minimization, selection and prioritization \cite{yoo2012regression}. Indeed, in the particular case of the Erlang language, most of the works in the area are focused on this specific task \cite{bozo2011selecting,taylor2012using,toth2010impact,toth2013reduction}. We can find other works in Erlang that share similar goals but more focused on checking whether applying a refactoring rule will yield to a semantics-preserving new code \cite{jumpertz2010using,li2007testing}.

With respect to tracing, there are multiple approximation similar to ours. In its standard libraries, Erlang implements two tracing modules. Both are able to trace the send and received messages, the function calls, and the process related events. One of these modules is oriented to trace the processes of a single Erlang node \cite{dbgErlang}, allowing for the definition of filters to function calls, e.g., with names of the function to be traced. The second module is oriented to distributed system tracing \cite{ttbErlang} and the output trace of all the nodes can be formatted in many different ways. Cronqvist \cite{redbug} presented a tool named redbug where a call stack trace is added to the function call tracing, making possible to trace both result and call stack. Till \cite{erlyberly} implemented erlyberly, a debugging  tool with a Java GUI able to trace the previously defined features (calls, messages, etc.) but also giving the possibility to add breakpoints and trace other features such as throw exceptions or incomplete calls. All these tools are accurate to trace specific features of the program, but none of them is able to trace the value of an arbitrary point of the program. In our approach, we can trace both the already defined features and also a point of the program regardless of its position.

\section{Conclusions}
\label{sec:conclusions}

During the lifecycle of any piece of software different releases may appear, e.g., to correct bugs, to extend the functionality, or to improve the performance.
It is of extreme importance to ensure that 
every new release preserves the correct behaviour of previous releases. Unfortunately, this task is often expensive and time-consuming, because it implies the definition of test cases that must account for the changes introduced in the new release. 

In this work, we propose a new approach to automatically check whether the behaviour of certain functionality is preserved among different versions of a module. The approach allows the user to specify a POI that indicates the specific parts of the code that are suspicious or susceptible of presenting discrepancies. Because the POI can be executed several times with a test case, we store the values that the POI takes during the execution. Thus, we can compare all actual evaluations of the POI for each test case. 
 
The technique introduces a new tracing process that allows us to place the POI in patterns, guards, or expressions. For the test case generation, instead of reinventing the wheel, we orchestrate a sophisticated combination of existing tools like CutEr, TypEr, and PropEr. But, we also improve the result produced by the combination of these tools introducing mutation techniques that allow us to find the most representative test cases. All the ideas presented have been implemented and made publicly available in a tool called SecEr.

There are several interesting evolutions of this work. One of them is to adapt the current approach to make it able to compare modules when some source of indeterminism is present (e.g., concurrency). We could also increase the information stored in traces with, e.g., computation steps or any other relevant information, so that we could also check the preservation (or even the improvement) of non-funtional properties such as efficiency. 
Another way to increase the usefulness of our tool is to permit the specification of a list of POI instead of a single one. 
This would allow us to trace several functionalities in a single run, or to strengthen the quality of the test suite. 
Finally, the integration with existing tools like Travis CI or similar would be very beneficial for the potential users. 

\bibliography{\biblio{biblio}}
\bibliographystyle{abbrv}

\end{document}

\newpage
\noindent \underline{Note for the reviewers:} The following appendix has been only included to ease the reviewing process, and it will not be part of the final paper. In case of acceptance, this appendix will be published as a technical report so that the interested reader will have public access to it.

\appendix
\label{appendix}


\section{Proofs of Technical Results}
\label{sec_proofs}
Pruebas.

\end{document}